\documentstyle[psfig,amstex]{mn}

\def\etal{et~al.}
\def\spose#1{\hbox to 0pt{#1\hss}}
\def\lta{\mathrel{\spose{\lower 3pt\hbox{$\mathchar"218$}}
     \raise 2.0pt\hbox{$\mathchar"13C$}}}
\def\gta{\mathrel{\spose{\lower 3pt\hbox{$\mathchar"218$}}
     \raise 2.0pt\hbox{$\mathchar"13E$}}}

\title[Dust in 3C324]{Dust in 3C324}

\author[P.~N.~Best \etal]{P.~N.~Best,$^1$ H.~J.~A.~R\"ottgering,$^1$
M.~N.~Bremer,$^2$ A.~Cimatti,$^3$ K.-H.~Mack,$^{4,5}$ \\
\\
{\LARGE G.~K.~Miley,$^1$ L.~Pentericci,$^1$ R.~P.~J.~Tilanus$^6$ and
P.~P.~van der Werf$^1$} \\ 
\\
$^1$ Sterrewacht Leiden, Postbus 9513, 2300 RA Leiden, The Netherlands\\
$^2$ Institut d'Astrophysique de Paris, 98bis Boulevard Arago, F-75014
Paris, France\\
$^3$ Osservatorio Astrofisico di Arcetri, Largo E. Fermi 5, I-50125
Firenze, Italy\\
$^4$ Istituto di Radioastronomia del CNR, Via P. Gobetti 101, I-40129
Bologna, Italy \\ 
$^5$ Radioastronomisches Institut der Universit\"at Bonn, Auf dem H\"ugel
71, D-53121 Bonn, Germany \\
$^6$ Joint Astronomy Centre, 660 North A'oh{$\overline o$}k{$\overline u$}
Place, University Park, Hilo, HI\,96720, US
}

\pagerange{\pageref{firstpage}--\pageref{lastpage}}
\pubyear{1998}
\begin{document}
\label{firstpage}

\maketitle

\begin{abstract}
\noindent The results of a deep submillimetre observation using SCUBA of
the powerful radio galaxy 3C324, at redshift $z=1.206$, are presented. At
850 microns, emission from the location of the host radio galaxy is
marginally detected at the 4.2$\sigma$ level, $3.01 \pm 0.72$\,mJy, but
there is no detection of emission at 450 microns to a 3$\sigma$ limit of
21\,mJy. A new 32\,GHz radio observation using the Effelsberg 100\,m
telescope confirms that the sub-millimetre signal is not associated with
synchrotron emission. These observations indicate that both the mass of
warm dust within 3C324, and the star formation rate, lie up to an order
of magnitude below the values recently determined for radio galaxies at $z
\sim 3-4$. The results are compared with dust masses and star formation
rates derived in other ways for 3C324.
\end{abstract}

\begin{keywords}
Galaxies: active --- Galaxies: individual: 3C324 --- Radio continuum:
galaxies --- Galaxies: ISM --- ISM: dust
\end{keywords}

\section{Introduction}

Searches for emission from distant galaxies at millimetre and
submillimetre wavelengths provide a direct method of investigating the
star formation activity of these galaxies, since a large proportion of the
optical and ultraviolet luminosity is reprocessed by dust to these longer
wavelengths (e.g. Soifer \& Neugebauer 1991).\nocite{soi91} In addition,
the strong negative K--correction caused by the steepness of the
Rayleigh--Jeans tail of the thermal dust spectrum means that at
submillimetre wavelengths equivalent objects will appear as bright at
redshift $z \sim 10$ as they do at $z \sim 1$ \cite{bla93}. Powerful radio
galaxies can be observed out to extreme redshifts, $z > 4$ (e.g. Rawlings
\etal\ 1996)\nocite{raw96a}, and therefore provide a probe through which
the star formation history of massive galaxies can be investigated.

The optical and ultraviolet morphologies of powerful radio galaxies with
redshifts $z \gta 0.6$ are dominated by emission elongated and aligned
along the direction of the radio axis \cite{mcc87,cha87}. The precise
details of this so-called `alignment effect' remain unclear, with
scattered quasar light, nebular continuum emission and jet--induced star
formation all likely to contribute at some level (e.g. see R\"ottgering \&
Miley 1996).\nocite{rot96d} Underlying this aligned emission, however, the
host galaxies of radio sources with redshifts $z \sim 1$ are well--formed
ellipticals whose stellar mass is dominated by an old stellar population
(e.g. Stockton \etal\ 1995)\nocite{sto95}, and which have settled down
into a stable $r^{1/4}$ law profile \cite{bes98d}.  These galaxies are
amongst the most massive at their epoch, with stellar masses of a few
times $10^{11} M_{\odot}$, and often appear to lie at the centre of young
or forming clusters \cite{dic97a,bes98d}.  Although a proportion of the
aligned emission may be associated with young stars, forming at a rate of
order $100 M_{\odot}$\,yr$^{-1}$ (e.g. Best \etal\ 1996, 1997a, Cimatti
\etal\ 1996), \nocite{bes96a,bes97b,cim96} the few galaxies that have been
studied in detail so far show no strong signatures in the rest--frame
ultraviolet\,/\,optical spectra that would indicate star--formation at
significantly higher rates (Cimatti \etal\ 1996, 1997).
\nocite{cim96,cim97}

At redshifts $z \gta 2$, powerful radio galaxies have much more clumpy
optical morphologies, with extreme examples showing many separate bright
emission regions within a spatial extent of 50 to 100\,kpc \cite{pen98}.
The individual emission clumps have sizes of 2 to 10\,kpc, and their
profiles and colours are similar to those of the UV--dropout galaxies at
$z \sim 3$ \cite{gia96b}. Although these radio galaxies overall display a
strong alignment effect, a significant proportion of the individual clumps
lie away from the radio axis, suggesting that their blue colours are not
directly associated with the alignment effect, but that these are more
likely to be star--forming objects, each with star formation rates of
order 10 solar masses per year.

At redshifts $z \gta 3$, targeted millimetre and submillimetre
observations have succeeded in detecting a number of powerful radio
galaxies (see Hughes \etal\ 1997 for a review, Ivison \etal\ 1998, Cimatti
\etal\ 1998, R\"ottgering \etal\ 1998, Hughes \& Dunlop 1998).
\nocite{hug97,ivi98a,cim98a,rot98b,hug98} The masses of warm dust implied by
these observations are a few times $10^8 M_{\odot}$ and the corresponding
star formation rates are extreme. If it is assumed that all of the dust is
heated by young stars, then thousands of solar masses per year of star
formation must be on-going. These results are frequently interpreted as
being the burst of star formation which forms the bulk of the stellar mass
of these galaxies. Supporting evidence for this hypothesis comes from a
deep Keck spectrum of the radio galaxy 4C41.17 ($z=3.8$) at rest--frame
ultraviolet wavelengths, which shows strong absorption features indicating
that the ultraviolet continuum of this galaxy is dominated by hot young
stars forming at rates of up to $1100 M_{\odot}$\,yr$^{-1}$ \cite{dey97}.

Despite the successes at the highest redshifts, submillimetre observations
of powerful radio galaxies at redshifts $z \sim 1$ have been sparse and
unsuccessful (e.g. see Hughes \& Dunlop 1998).\nocite{hug98} Studies of these
galaxies are of great importance, firstly for understanding the continued
evolution of the interstellar medium of massive galaxies, and secondly for
investigating the extent to which the powerful active galactic nucleus
(AGN) can be responsible for heating the dust within the radio
galaxies. The advent of highly sensitive instruments such as the
Submillimetre Common--User Bolometer Array (SCUBA; Holland \etal\
1998)\nocite{hol98} on the James Clerk Maxwell Telescope\footnote{The JCMT
is operated by the observatories on behalf of the UK Particle Physics and
Astronomy Council, the Netherlands Organisation for Scientific Research
(NWO) and the Canadian National Research Council.}  (JCMT) has opened
up the possibility of more detailed studies of these objects.

In this letter we present submillimetre observations of one such radio
galaxy, 3C324 at redshift $z = 1.206$. The very strong alignment effect
and the scattering properties of this galaxy, together with a central
absorption feature in the HST image, make it one of the most promising
candidates in the 3CR $z \sim 1$ sample for a large dust mass
\cite{bes96a,cim96}. In Section~\ref{observ} we describe the observations
of 3C324 and the data reduction. In Section~\ref{results} we present our
results and discuss their implications.  Throughout this letter we adopt
$q_0$=0.5 and $H_0 = 50$\,km\,s$^{-1}$Mpc$^{-1}$.
 
\section{Observations}
\label{observ} 

\subsection{SCUBA observations}

3C324 was observed using SCUBA on the JCMT for a total of 105 minutes on
source during the nights of November 30th and December 2nd, 1997. SCUBA
has two arrays of bolometric detectors, cooled to 0.1K. The
short--wavelength (SW) array has 91 feedhorns and can be operated at 350
or 450\,$\mu$m; the diffraction--limited beamwidth at 450\,$\mu$m has a
half--power diameter of 7.5 arcsec. The long-wavelength (LW) array has 37
feedhorns and can be operated at 750 or 850\,$\mu$m; at 850\,$\mu$m the
beamwidth is 14 arcsec. A dichroic beam--splitter enables observations
using both arrays simultaneously.

In the standard SCUBA photometry mode, observations are made using the
central pixels of each array, which are aligned with each other to within
about an arcsecond. A jiggling procedure is employed whereby the secondary
mirror is jiggled around a 9--point filled square, with a 2 arcsec offset
between each pointing. This method has been shown to produce the best
photometric accuracy (cf Ivison \etal\ 1998)\nocite{ivi98a}. The
integration time for each pointing of the jiggle is 1 second. During this
period the secondary mirror is chopped at 7 Hz between the source and a
reference sky position, usually with a chop-throw of between 60 and 180
arcsec in azimuth. Following the 9 second jiggle, the telescope is nodded
so that the chop position is placed at the opposite side of the source and
the jiggle pattern is repeated.

The observations of 3C324 were made at 850\,$\mu$m and 450\,$\mu$m using
the two--bolometer photometry mode, which is an adaptation of the standard
photometry mode. In this mode the chop--throw is not fixed in azimuth, but
is fixed such that, in the primary nodding position, the source is chopped
to the centre of one of the other bolometers of the LW array. This `chop
bolometer' is chosen to be one for which the chop--throw lies as nearly as
possible in azimuth, and whose position corresponds closely to that of a
bolometer in the SW array. In two--bolometer mode, the source is thus
observed for half of the total integration time in the central bolometer,
plus a further quarter in the chop bolometer, improving the observing
efficiency.  Due to the apparent curvature of the arrays on the sky, the
chop--throw in the reference nodding position does not centre the source
on a bolometer. A three--bolometer photometry mode would maximise the
observing efficiency and is planned for SCUBA, but was not operational at
the time of these observations.

350 18--second integrations were made of 3C324, split into 7 sets of 50
integrations. Telescope pointing checks were made using 1611+343 before
each set of integrations; the offsets were small, typically below 2
arcsecs. Skydips were made at intervals of about 2 hours throughout both
nights, and showed the sky to be stable. On November 30th the atmospheric
zenith opacity was consistently about 0.12 at 850\,$\mu$m all night and
between 0.55 and 0.65 at 450\,$\mu$m. On December 2nd the opacity was
about 0.25 at 850\,$\mu$m and 1.6$\pm$0.3 at 450\,$\mu$m. The observations
were made whilst 3C324 was at low airmass (mean value 1.24). Calibration
observations, using 3C273 on the first night and IRC+10216 on the second,
were taken using the same two--bolometer photometry mode, the fluxes of
these objects being bootstrapped to Mars. The calibration uncertainty for
these data is estimated to be $\lta 10$\% at 850$\mu$m, but possibly as
much as 25\% at 450$\mu$m.

The data were reduced using the SCUBA software, SURF \cite{jen98}. The
reference measurements were subtracted from the signal beams, the
bolometers were flatfielded using the standard SCUBA flatfield, and the
extinction correction was applied. Noisy integrations and strong spikes in
individual bolometers were rejected; this removed a little under 5\% of
the data. The residual sky background was removed by subtracting the mean
signal from the off--source bolometers in the inner ring around the
central bolometer, excluding those bolometers which had significantly
higher than average noise (see also Ivison \etal\ 1998).\nocite{ivi98a}

The integrations were then concatenated to form a single dataset, and the
consistency of this dataset was investigated using a two sample
Kolmogorov--Smirnov (KS) test. In this way, periods when the changes in,
for example, the atmospheric conditions or the telescope focus may have
affected the data quality can be identified and removed \cite{jen98}.  The
data was split into subsamples, and the first two samples were compared
for consistency. The second sample was rejected if the probability that
the two samples are drawn from the same parent sample was below a given
limit; otherwise, it was concatenated with the first sample. The resulting
sample was then compared with the third subsample, the process was
repeated, and so on. For each wavelength the KS test was run on both the
central bolometer and the chop bolometer with a variety of inputs: the
number of subsamples, the order in which they were supplied, and the limit
for rejecting the samples were all varied. The results obtained were
consistent, and rejected about 5\% of the data.

At 850\,$\mu$m, the signal measured in the central bolometer was $3.65 \pm
1.17$\,mJy, and that in the chop bolometer $4.46 \pm 2.55$\,mJy (the error
in the latter measurement is higher since this bolometer was only
on--source for half as long as the central bolometer). Combined, these
give $3.78 \pm 1.05$\,mJy, a signal with a 3.6$\sigma$ significance. 3C324
has also been observed using SCUBA by Hughes \& Dunlop \shortcite{hug98}.
They determine a flux density at 850\,$\mu$m of $2.4 \pm 1.0$\,mJy.
Combining this with our data gives a combined signal of $3.01 \pm
0.72$\,mJy, a marginal detection at the 4.2$\sigma$ level.  At 450\,$\mu$m
no signal was detected to a 3$\sigma$ upper limit of 21\,mJy.

\subsection{Effelsberg Observations}

3C324 was observed using the 100\,m Effelsberg Telescope on July 15th
1998, at a central frequency of 32\,GHz. The 3-feed receiver system
installed in the secondary focus was used in a multi-beam mode. Each horn
feeds a 2-channel receiver with an IF polarimeter providing full Stokes
information simultaneously. The bandwidth was 2 GHz.

The observations were made using a cross-scan in the equatorial coordinate
system, with the main beam scanning a distance of 4 arcmin at a scanning
speed of 10 arcsec per second. The offset feeds were used to efficiently
remove atmospheric noise. 32 such sub--scans were made of 3C324 (16 in a
north--south direction, and 16 east--west), and the data in the combined
scan was sampled at 3.8 arcsec intervals.

4 sub--scans of the source 3C286 were also made, and used to calibrate the
flux density of 3C324, taking the flux density of 3C286 at 32\,GHz to be
1.779\,Jy according to the scale of Ott \etal\ \shortcite{ott94}, which
corresponds to the Baars scale \shortcite{baa77} at lower frequencies. The
flux density of 3C324 at 32\,GHz was thus determined to be $32.4 \pm
9$\,mJy.

\section{Discussion}
\label{results}

\begin{figure*}
\centerline{
\psfig{figure=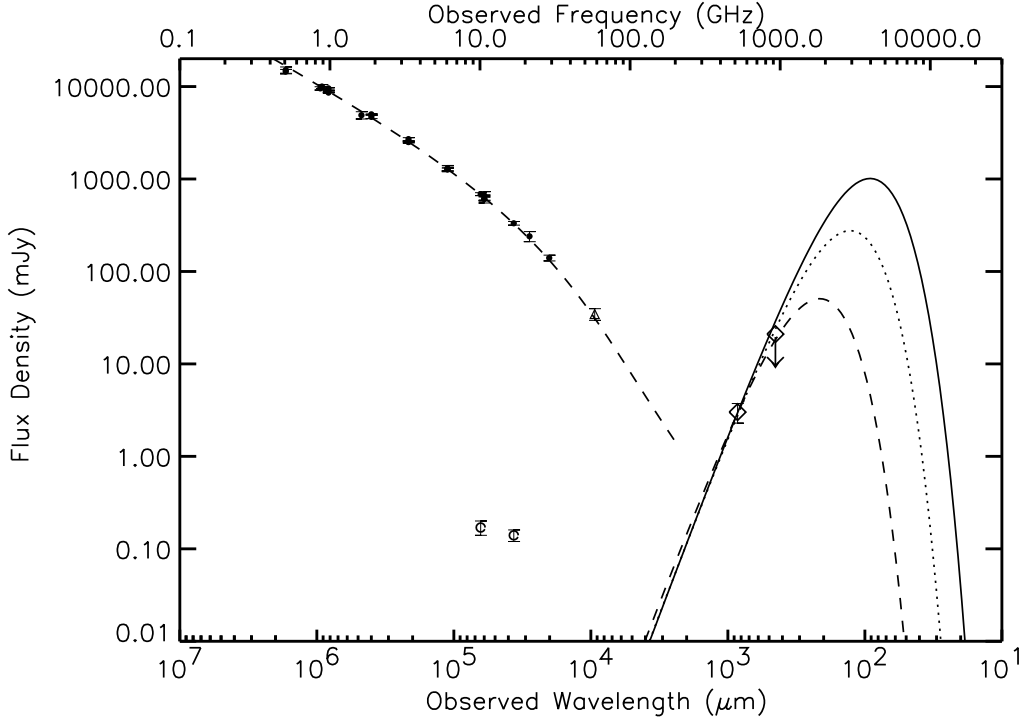,clip=,width=13.6cm}
}
\caption{\label{scubfig} The spectral energy distribution of 3C324. The
open diamonds represent the current SCUBA data, and the solid, dotted and
dashed lines represent isothermal grey--body emission with an emissivity
index $\beta = 2$ for dust at temperatures of 70K, 50K and 30K
respectively. The filled circles are the radio flux densities of 3C324,
taken from Herbig \& Readhead (1992) and Best \etal\ (1998a). The open
triangle is our new 32\,GHz data point. The dash--dot line shows a single
power--law fit to the synchrotron emission at radio frequencies below
10\,GHz. The open circles are the flux densities of the radio core
emission only, taken from Best \etal\ (1998a). }
\end{figure*}

In Figure~\ref{scubfig} we plot the results of these SCUBA observations,
together with the flux densities of the synchrotron emission at radio
wavelengths from the literature (using the compilation of Herbig \&
Readhead 1992, and new 5 and 8 GHz measurements from Best \etal\
1998a)\nocite{her92}, and our new Effelsberg data.

The radio synchrotron emission of double radio sources steepens at high
frequencies due to electron cooling, the precise details of the steepening
depending upon the synchrotron ageing model adopted. The radio spectrum of
3C324 was compared with the three popular ageing models, and the best fit
was provided using a KP model \cite{kad62,pac70}.  This is shown on
Figure~\ref{scubfig}: the fitted synchrotron spectrum passes more a factor
of 10 below the 850\,$\mu$m SCUBA signal. It is beyond the scope of this
letter to discuss in detail the synchrotron ageing models: we merely note
that although the KP model reflects a somewhat unphysical situation, since
it does not allow a uniform distribution of pitch angles, Carilli \etal\
\shortcite{car91} showed that this model also provides the best match to
the lobe emission of Cygnus A, a low redshift analogue of 3C324;
therefore, it provides an acceptable template.

The hotspots and radio core, where the electrons are continuously injected,
may, however, show less spectral steepening. The radio core flux densities
at 5\,GHz and 8\,GHz are only 0.17 and 0.14\,mJy respectively (Best \etal\
1998a; see Figure~\ref{scubfig})\nocite{bes98a}, and so its flux density
at 850\,$\mu$m will be negligible. The hotspots, which lie towards the
extremes of the SCUBA 850 micron beam (the radio source angular extent is
11.5 arcsec, compared to a beam FWHM of 14 arcsec), cannot be resolved
from the lobe emission even in our high resolution 8\,GHz maps, and so
precisely determining their contribution to the radio source flux density
is not possible. The poor agreement of a continuous injection synchrotron
model with the 15 and 32\,GHz data points, however, suggests that they do
not dominate the flux density even at the highest radio frequencies. In
conclusion, the shape of the high frequency radio spectrum appears to rule
out the high frequency tail of the synchrotron emission as a possible
origin of the 850\,$\mu$m signal.

\nocite{bes98a}

Although we cannot categorically state that a signal with 4.2$\sigma$
significance constitutes a detection, the agreement between the signal we
derived and that derived from an independently data set by Hughes and
Dunlop \shortcite{hug98}, and the consistency between the two bolometers and
stability of the signal throughout the seven sets of 50 integrations in
our observations (a positive signal was measured in ten of the fourteen
measurements), adds support to its reality. For the remainder of this
letter we shall perform calculations based upon an 850\,$\mu$m flux
density of 3.01\,mJy: it should be realised, however, that the values
derived should strictly be treated as upper limits instead of derived
quantities.

The mass of warm dust, $M_{\rm d}$, can be calculated from the
sub-millimetre flux using the equation:

\begin{displaymath}
M_{\rm d} = \frac{S(\nu_{\rm obs}) D_{\rm L}^2}
{(1+z)\kappa_{\rm d}(\nu_{\rm rest})B(\nu_{\rm rest},T_{\rm d})}
\end{displaymath}

\noindent where $S$ is the observed flux density, $\nu_{\rm obs}$ and
$\nu_{\rm rest}$ are the observed and rest--frame frequencies, $D_{\rm L}$
is the luminosity distance, $z$ is the redshift, $\kappa_{\rm d}$ is the
mass absorption coefficient, $B$ is the black--body Planck function, and
$T_{\rm d}$ is the dust grain temperature. To allow comparison with the
discussion of the $z \gta 3$ radio galaxies \cite{hug97}, we adopt a dust
temperature $T_{\rm d} =50$\,K, and a mass absorption coefficient
$\kappa_{\rm d} = 0.067 (\nu_{\rm rest} / 250 {\rm
GHz})^{\beta}$\,m$^2$\,kg$^{-1}$ with $\beta=2$. This provides a mass of
warm dust in 3C324 of $1.2 \times 10^8 M_{\odot}$. If it is assumed that
this dust is heated primarily by young stars, then a simple scaling
between the star--formation rate and the submillimetre luminosity can be
obtained using nearby starbursts such as M82 \cite{hug94}, providing
a current star formation rate for 3C324 of $350 M_{\odot}$yr$^{-1}$.

Adoption of a lower temperature (the 450\,$\mu$m SCUBA upper limit implies
that if the 850\,$\mu$m is real then the dust is at a temperature $T \lta
45$\,K) would increase the dust mass, by about a factor of 2 for $T_{\rm
d} =30$\,K. The dust mass would also increase if a flatter grain
emissivity index were used (about a 50\% increase for $\beta = 1.5$). A
full discussion of the uncertainties of these values can be found in
Hughes \etal\ \shortcite{hug97}. The dust masses derived are also strongly
dependent upon $H_0$ and $q_0$ through the luminosity distance dependence:
for $H_0 = 100$\,km\,s$^{-1}$\,Mpc$^{-1}$ the mass would be a factor of 4
lower than that derived here; for $q_0 = 0$ it would be nearly a factor of
2 higher. All these conversion factors would, however, apply similarly to
any star formation rates determined from the optical and ultraviolet
emission, and (in the case of $q_0$, to an even greater extent) to the
dust masses derived for the highest redshift radio galaxies.

The derived mass of warm dust within 3C324, $1.2 \times 10^8 M_{\odot}$,
can be compared to dust masses derived in other ways for this radio
galaxy.  HST images of the galaxy show bright extended ultraviolet
emission, but the central regions of the galaxy are obscured, probably by
a dust lane with $E(B-V) \gta 0.3$ \cite{lon95,dic96}. The mass of this
centrally concentrated dust can be related to the extinction using the
equation $M_{\rm d} = \Sigma \langle A_{\rm B}\rangle / \Gamma_{\rm B}$,
where $\Sigma$ is the area covered by the dust extinction, $\langle A_{\rm
B}\rangle \approx 4 E(B-V)$ is the mean B--band extinction, and
$\Gamma_{\rm B} \approx 8 \times 10^{-6}$\,mag\,kpc$^{2} M_{\odot}^{-1}$
is the B--band mass absorption coefficient \cite{sad85}. The dust
extinction in the central regions of 3C324 covers an area of at least 2 by
2\,kpc, requiring a minimum of $\sim 10^6 M_{\odot}$ of dust to be
concentrated near the nucleus of the radio galaxy.

The extended emission of 3C324 is strongly polarised due to scattering of
radiation from a hidden quasar nucleus, and the shape of the polarised
flux spectrum indicates that, regardless of whether the scattering medium
is electrons or dust, the scattered light must be dust reddened. In the
case of pure dust scattering, a mass of at least a few times $10^6
M_{\odot}$ of dust must be distributed throughout the galaxy
\cite{cim96}. In addition to scattering the AGN emission, this same dust
would absorb some of the AGN radiation and reprocess this to submillimetre
wavelengths.  These two lower limits allow the
mass of warm dust in 3C324 to be constrained to within an order of
magnitude even if the sub-millimetre derived value is considered as an
upper limit.

The star formation rate of $350 M_{\odot}$\,yr$^{-1}$ deduced for 3C324
from the SCUBA observations can be compared with a star--formation rate
estimated from the rest--frame 2800\AA\ flux density. The Keck spectrum of
Cimatti \etal\ \shortcite{cim96} provides a 2800\AA\ continuum flux
density of $2.6 \times 10^{-18}$\,erg\,s$^{-1}$cm$^{-2}$\AA$^{-1}$ from a
3.8 by 1 arcsecond slit along the galaxy. Extinction by a few times $10^6
M_{\odot}$ of dust spread evenly throughout this region would mean that
the intrinsic flux would be about 50\% higher. However, the strong
UV/optical polarisation of this galaxy suggests that between 30 and 50\%
of the 2800\AA\ flux density of 3C324 is associated with a scattered
component \cite{cim96}, and the nebular continuum contribution will also
be highly significant, so only a fraction $0 < f \lta 0.3$ will be
associated with young stars.  Comparison with the results of Madau \etal\
\shortcite{mad98} for the 2800\AA\ luminosity expected for on--going star
formation (for a Scalo \shortcite{sca86} IMF with upper and lower
cut--offs of 125 and 0.1\,$M_{\odot}$ respectively, and solar
metallicity), indicates that the rate of on--going star formation must be
below $f \times 130 M_{\odot}$\,yr$^{-1}$. If, on the other hand, the star
formation were associated with only a short burst induced by the radio
source, about $5 \times 10^6$\,yr old, the Bruzual \& Charlot
\shortcite{bru93} stellar synthesis codes show that the measured 2800\AA\
flux density corresponds to a star formation rate of $f \times 1050
M_{\odot}$\,yr$^{-1}$, just consistent with the value derived from the
current submillimetre observations if $f$ lies at the upper end of its
allowed range.

With these star formation rate limits, powerful radio galaxies at redshift
one cannot be undergoing anything more extreme than short--lived ($\lta
10^7$ yr) starbursts of a few hundred solar masses per year. This is in
stark contrast to the situation in radio galaxies at redshifts $z \gta 3$
where star--formation rates of several thousands of solar masses are
derived. These values are based upon the assumption that the dust is
heated by young stars rather than by the AGN. The current observations
support that hypothesis: the AGN of the high redshift galaxies have a
similar intrinsic power to that of 3C324, whilst the submillimetre
emission from 3C324 at least a factor of a few lower. Therefore, only a
small fraction of the dust in the very distant galaxies can be heated by
the AGN.
\medskip

\noindent {\it Note added in manuscript:} On August 19 1998, a deeper (64
sub-scans) Effelsberg 32\,GHz observation of 3C324 was made during better
weather conditions, using the same observational set-up as described in
the text. An improved flux density measurement of $38.5 \pm 3.6$\,mJy was
determined, strengthening the conclusion of a turn-over in the
synchrotron spectrum.

\section*{Acknowledgements} 

This work was supported in part by the Formation and Evolution of Galaxies
network set up by the European Commission under contract ERB
FMRX--CT96--086 of its TMR programme. HJAR acknowledges support from a
programme subsidy granted by NWO and a NATO research grant. KHM was
supported by the European Commission TMR Programme, Research Network
Contract ERB FMRX--CT96--0034, `{\it Ceres}'. The authors thank Goran Sandell
for useful discussions about calibration of SCUBA data, and Helge Rottmann
for carrying out the Effelsberg observation.

\label{lastpage}
\bibliography{pnb} 
\bibliographystyle{mn} 

\end{document}